\documentclass[conference]{IEEEtran} 
\usepackage{enumerate}
\usepackage{tablefootnote}
\usepackage{url}
\usepackage{balance} 

\begin{document}

\hyphenation{Oka-nagan pro-ject pro-jects Sou-rce cour-ses star-ted wi-de-ly com-mu-ni-ca-ti-on syn-chro-ni-za-ti-on}


\title{Parallel Programming Applied Research Projects for Teaching Parallel Programming to Beginner  Students}



\author{\IEEEauthorblockN{Youry Khmelevsky}
\IEEEauthorblockA{Computer Science, Okanagan College\\ Kelowna, BC, Canada\\
Email: ykhmelevsky@okanagan.bc.ca}\\
\IEEEauthorrefmark{1}{Also Affiliated with Mathematics, Statistics, Physics, and Computer Science}\\ {Irving K. Barber School of Arts and Sciences, UBC Okanagan, BC Canada}
\and
\IEEEauthorblockN{Ga\'etan J.D.R. Hains}
\IEEEauthorblockA{LACL, Univ. Paris-Est Cr\'eteil\\ Paris, France \\
Email: gaetan.hains@gmail.com}
}

\maketitle 
  
\begin{abstract} 
In this paper, we discuss the educational value of a few mid-size and one large applied research 
projects at the Computer Science Department of Okanagan College (OC) and at the Universities of Paris 
East Cr\'eteil (LACL) and Orl\'eans (LIFO) in France. 

We found, that some freshmen students are very active and eager to be involved in applied 
research projects starting from the second semester. They are actively participating in programming 
competitions and want to be involved in applied research projects to compete with sophomore and 
older students. Our observation is based on five NSERC Engage College and Applied Research and 
Development (ARD) grants, and several small applied projects. 

Student involvement in applied research is a key motivation and success factor in our activities,
but we are also involved in transferring some results of applied research, namely programming techniques, 
into the parallel programming courses for beginners at the senior- and first-year MSc levels. 
We illustrate this feedback process with programming notions for beginners, practical tools to acquire them 
and the overall success/failure of students as experienced for more than 10 years in our French University courses. 
\end{abstract}

%
%

%
%

%
%



\section{Introduction}
This paper is an attempt to analyze our results of 12 years of small and mid-size applied research projects introduction to students initially started at COSC 224 Capstone Project course (4th semester) and at COSC 470/471 Software Engineering Capstone Projects at the Computer Science department at Okanagan College and then to freshmen students from 2nd semester in sponsored by NSERC and by industry applied research projects. We also analyze the results of almost two decades of teaching beginner's parallel programming at the Universities of Orl\'eans and Paris-Est Cr\'eteil, in this case to final-year undergraduates and first-year MSc Computer Science students \cite{khmelevsky2007information}, \cite{govorov2005security}, \cite{
Khmelevsky:2009:SDP:1536274.1536292}, \cite{khmelevsky2009okanagan}. Every year, starting from 2005 we searched for new solutions and new pedagogical approaches within COSC 224 --- Projects in Computer Science, COSC 470 --- Software Engineering and COSC 471 --- Software Engineering Project Project courses \cite{khmelevsky2010cloud}, \cite{Khmelevsky:2011:DLC:1989622.1989627},  \cite{Khmelevsky:2011:ICS:1989622.1989637},  \cite{Khmelevsky:2011:RTS:1989622.1989638}, \cite{5999857}. Starting from 2007 we tried to introduce research projects to all our capstone projects and some local industrial 
clients supported our idea \cite{Khmelevsky:2016:TYC:2910925.2910949}, but it took almost 7 extra years to get support for our student research or research related projects from NSERC. Several industrial projects were so successful that we were able to obtain an NSERC Community College Innovation (CCI) Applied Research grants (Engage College grants), two Mitacs grants with SFU and UBCO, an Amazon's AWS in Education Research grant, and several small grants funded by the College for the student research projects with industrial sponsors \cite{Khmelevsky:2012:ACG:2247569.2247578}, \cite{khmelevsky2013strategies}, \cite{Trevor2014}, \cite{khmelevsky2015hybrid}. 

As other educators and computer science (CS) researchers discovered, CS ``research has an opportunity to change all 
other disciplines. Much of the current CS research is of an incremental {\em stovepipe} or inward-looking nature, even 
though CS has a unique ability to fundamentally transform virtually every other discipline" \cite{
Feamster:2008:GRT:1352322.1352294}. Moreover,  the ``undergraduate research experiences are widely promoted in science, 
technology, engineering, and mathematics (STEM) for their potential to bring several benefits to students, like: (1) 
improved retention in both the major and discipline-related careers;  (2) ability to work independently and to communicate 
well with a team; (3) increased confidence in academic knowledge and technical skill; (3) broader awareness of the 
discipline and and awareness of career opportunities and support for making career choices."

Another researcher in \cite{Barker:2009:SFP:1513593.1513598} concluded:  ``these positive outcomes are not guaranteed simply 
because a student participates; however, the experience must truly be an experience of research and with enough structure 
that the student is able to be successful." 

A few years ago several freshmen students contacted us with requests to organize students programming competitions and 
to be accepted into our undergraduate research projects. They successfully competed in ICPC and IEEExtreme programming 
competitions and achieved good results. As an experiment, we accepted a few freshmen students to our research projects 
and we found, that some talented freshman students can contribute to the research projects on the same 
level, if not better as 3rd and 4th years undergraduate students. 

In the next section we will discuss our ongoing projects related to automatic code generation and parallel programming 
programming initial project results. This is concrete evidence of the upward transfer of skills from undergraduate 
education to applied research in parallel programming. We also summarize results of a long-term experiment in 
teaching parallel programming to beginners with simplified and formally verified programming interfaces, that are 
the result of applied research projects by academics and PhD students. 

\section{Teaching Parallel Programming for Freshmen Student in Applied Research Projects} About 7 years ago we started 
parallel programming and automatic code generation as small applied research projects within our capstone framework at the Computer 
Science Department of Okanagan College, and also at the UBC Okanagan Campus , \cite{7116828}\cite{Khmelevsky:2013:STT:2555523.2555540}, \cite{7237130},  \cite{ouimet2016game}. 
Students enjoyed projects within capstone project-courses \cite{7116828}, \cite{mcdonald2016biometrie}, \cite{mcdonald2016sport}, \cite{Khmelevsky:2016:NPT:2910925.2910937}  and 
several projects resulted in student research papers  \cite{atkinson2016reporting}, \cite{hains2016game},  \cite{7740432}, \cite{7726818}. 
Starting from 2016 we also accepted freshmen students to the parallel programming and automatic code generation small applied research projects \cite{hains2017game}, \cite{ward2017gaming}, \cite{cocar2017utilizing}, \cite{hains2019natural}, \cite{hains2020wtfast}, \cite{mazur2020machine}. We found that such 
projects, additionally to a small stipend for the student research work creates a strong motivation in courses starting 
from the first year. Students see immediate application of their knowledge in the small applied research 
projects at the College. They understand better the need for first year courses such as COSC 126---System 
Analysis and Design, COSC 109 --- Technical Aspects of Operating Systems, COSC 118 --- Networks and Telecommunications I, COSC 
131 --- Visual Programming, and two communication courses in addition to traditional programming courses.  

The second author (Hains) assisted with project goals development and with the students supervision. 
He taught for about 20 years bulk-synchronous parallel (BSP) programming, 
which is based on a paradigm of global synchronous communications allowing a realistic performance (``cost") model
---that leads to portable and scalable algorithm design, or in the case of beginner students, the realistic prototyping 
of parallel algorithms without parallel machine. 
We already introduced BSP parallel programming paradigm into small applied research projects successfully 
\cite{7084518} and continue our educational development. 

In 2017 we started a new freshman parallel programming short applied research project on the performance 
evaluation of the Parallel MATLAB library.  This is both an exercise in declarative numerical programming, and 
a systematic introduction to performance benchmarking and modelling. 
In previous research projects sophomore and upper level students successfully investigated servers workload 
and many different applications/game servers performance issues \cite{7084518}, \cite{7740432}, \cite{7726818}, 
\cite{7237149}, but those projects were too complex for the freshman students. 
After observing this problem we adapted our tasks and research projects for very young, but skillful students. 

We assigned the following tasks for the freshmen students realizing the parallel MATLAB project:
\begin{itemize}
  \item Install the Parallel MATLAB library.
  \item Try examples with it to write pure-declarative code (pure operations on arrays) and obtain multicore/multinode 
        parallel execution. 
  \item Design a simple performance model for it (how fast for how many cores, how much communication etc) and test that 
        model against measurements. The ``model" can be as simple as a low-degree polynomial giving execution time against 
		data dimensions and the number of processing cores. 
  \item Document all the test data with the experiment setup: software (SW) versions, hardware, number of cores used, 
       (number of threads if its visible), measured quantity (time, space etc). 
  \item For a simple operation like Matrix x vector that would mean the size of matrix etc show the Matlab expressions 
       what are compiling. 
  \item Then try to cover the space of tests to obtain ideally 3D graphs with (time or memory) vs cores x (data size). 
       From that we can interpolate surfaces (or just curves if the graphs were 2D) that with the help of your supervisor(s) 
	   guess from parallel performance models. That helps understand the general compiler or algorithm as applied to a 
	   specific parallel machine (in this case a one-node or two-node multicore PC). 
	   Each graph is about running the same Parallel MATLAB expression for various cores x (data size).
  \item Finally if time allows, a more creative project phase can lead to the initial phase of a serious project: 
        design a language-based general cost model that can unify all the previous performance measurements in a 
		syntax-based model of performance. 
		In other words, design a parallel cost model for a whole sub-language of Parallel MATLAB. 
\end{itemize}

The list of tasks above is just an initial investigation, but the freshman students were able to install MATLAB, add 
necessary libraries and run almost all tests in a few weeks. The project is still on-going, but students are excited and 
like to do experiments with the parallel code execution and coding experiments. Such a project is a good example of a 
simplified research project (namely, benchmarking and modelization of the a parallel library's performance) with 
quantitative objectives that can be converted into a precisely defined and progressively difficult student project. Given 
enough time and effort it could lead young undergraduates to Master's level research topics and an introduction to the 
general research problems of parallel programming and parallel software engineering. 

\section{Research results in parallel programming and MSc-level teaching} 

The BSP (bulk-synchronous parallel) model of parallel computing \cite{Valiant2008},  \cite{ McColl1993b}, \cite{hains2018algorithmes} has been studied, 
applied and implemented since the late 1990s to improve the reliability, predictability and {\em portable} scalability of 
parallel programs. It is based on the notion that a parallel algorithm is a sequence of {\em supersteps} where a vector of 
$p$ computation units (cores, nodes, computers) execute as many asynchronous and independent phases of computation (threads, 
processes) before all exchanging necessary data in a globally synchronous communication phase. The advantages of the BSP 
model over ad-hoc parallel methods are many: determinism, a performance model that relates the communication phase to its 
underlying architecture in a linear way, possible checkpoints or debugging steps at the end of supersteps, ``immortal" 
algorithms that can adapt to any future architecture etc. 

We have taught theoretical BSP algorithms to doctoral students since the invention of the model by Valiant ca. 1988 and its 
in-depth investigation by McColl in the 1990s. We then developed a declarative API for programming BSP algorithms that can 
serve both as a parallel functional language and simplified tool for learning parallel programming. This API is called 
BSMLlib or ``BSP-ML library" \cite{Bousdira2010} because it is based on the OCaml (Objective-Ca-ML) functional language, a 
dialect of the ML language (a strongly-typed form of Lisp). Whatever the host language and functional programming details, 
BSMLlib summarized BSP programming to the following primitives: 
\begin{itemize}
\item A global, load-time constant {\tt nprocs} for the number $p$ of processing units. 
\item A constructor and a transformer for building asynchronous computation steps of width $p$. 
      Those operators binds the local variable {\tt pid} that identifies a local thread by its rank $0,1,\ldots,p-1$. 
\item A destructor for folding parallel data into a local array or list that must then be processed locally and sequentially. 
\item A transformer based on a $p\times p$ relation of message sending-reception for creating global 
      communication-synchronization phases. 
\end{itemize} 

We have used BSMLlib to teach parallel programming to final-year undergraduates and first-year MSc students so that they 
learn the expected performance (BSP model) of any simple parallel program written with BSMLlib. The most demanding exercises 
involved advanced parallel data structures like balanced parallel hashing, or simulations like the n-body problem, or BSP 
sample-sort programming and benchmarking. One great advantage of BSMLlib is that a purely-sequential simulator of the 
library (1000 lines of OCaml) is able to produce exact counts of the communication and synchronization costs. If 
complemented with a bytecode-instruction counter for the asynchronous parallel phases, it is able to produce realistic and 
scalable estimates of runtime without any more complex technology than the bytecode-interpreted version of OCaml (which runs 
on Unix, Windows, a web-server service at OCamlPro SARL and even an Android port). 

In all classes we observed that a large portion of the students became familiar and proficient with the constructor-
destructor sub-language which allowed them to write a large subset of our elementary algorithms (sorting being a notable 
exception). The the communication-synchronization transformer (called {\em put}) requires a two-dimensional coding of 
messages which is objectively difficult to acquire. In a more recent research project, \cite{Li2012} we designed a 
simplified form of BSMLlib called SGL (scatter-gather language) where the transformer operator {\tt put} is absent so that 
all global operations appear like one-to-many constructors or many-to-one destructors. This language has the same 
advantages as BSMLlib plus: 
\begin{itemize}
\item Is easier to learn as experienced in class and for a very precise reason learned from BSMLlib teaching. 
\item A centralized structure than can be nested to simulate heterogeneous architectures that are not ``flat" 
      e.g. multi-node systems of multicore nodes. 
\item Can express more than 80\% of a set of basic application libraries (like list- and array APIs) without the use 
      of a $p\times p$ operator like {\em put}. 
\end{itemize} 
Again SGL, the result of parallel programming research, has been applied to improved teaching for the benefit of all 
beginners, at well as inspiring new concrete ideas in parallel software engineering. 

\section{CONCLUSION}
\label{Conclusion}
Software development skills are a key asset of modern economies. 
Training undergraduates in parallel programming is an important objective that has not yet been well 
covered by the CS curriculum. The following quotations from educators illustrate the general importance 
of teaching methods. ``The software development and software development project courses with external and 
internal sponsors drastically increased learning activity, creativity, productivity and success of student work. 
Students like to see computer science as social, relevant, important, and caring endeavour" 
\cite{Buckley:2009:VCS:1498765.1498779}.

``The research workforce of the future is important for many STEM disciplines, but the computer science research 
workforce has become critical to the US and global economies" \cite{Barker:2009:SFP:1513593.1513598} and to Canada. 
``All of the engineering and science discoveries and innovations are now dependent on computational science" 
\cite{Khmelevsky:2011:RTS:1989622.1989638}.  
``Students recognize that research is a collaborative experience requiring communication skills to complete 
and disseminate the work" \cite{Milani:2009:REU:1565799.1565821}. 

As a faculty at Okanagan College mentioned, we have a new generation of students, 
are looking for challenges during their study at post-secondary institutions. 
They like to work hard, but they want to be rewarded for their hard work. 
Many students apply to the College because they like small classes and the lower tuition to 
 compare with the universities. On the other hand they are demanding a high quality education,  involvement in the worldwide programming competitions (IEEEXtreme 24-Hour Programming Competition\footnote{\url{http://www.ieee.org/membership_services/membership/students/competitions/xtreme/index.html}}, the ACM International Collegiate Programming Contest (ICPC)\footnote{\url{https://icpc.baylor.edu/}}, and others), and  involvement in industrial applied research projects, especially supported by Government and/or by large corporations. Many freshmen students want to continue their education at top universities within MSc and even in PhD programs. We believe that the results of MSc-level teaching and research work can be applied in very concrete ways to help more undergraduates acquire top skills in the development and performance-modelling of parallel programs. In return, undergraduate teaching can help improve parallel programming tools themselves, as well as motivating some of the students to pursue their technical training in the form of applied research projects. \\
 
The experiences described here somewhat contradict the traditional dichotomy between on the one hand, community-college teaching to undergraduates who have no exposure to research, and on the other hand University  students who are exposed to the results of ongoing research projects. \\

We are confident that they are applicable to many other areas of software engineering,  if not most of the computing disciplines. 

\section{ACKNOWLEDGMENTS}
We thank the Natural Sciences and Engineering Research Council of Canada for supporting five our applied research projects 
in 2014--2017.  Our thanks go also to the AWS Programs for Research and Education program for supporting our research and educational projects, as well as to Atlassian. We thank the IBM Academic Initiative program for supporting us with Rational Rose engineering software, licenses and with training materials. 

Research on BSMLlib and SGL has benefited from many French-government grants including: the CARAML national project 
sponsored by the Ministry of Research, the HPIAF project sponsored by EXQIM S.A.S, Chong Li's CIFRE doctoral project 
sponsored by ANRT and EXQIM. The second author thanks his many partners in BSMLlib research: Fr\'ed\'eric Loulergue, 
Fr\'ed\'eric Gava and Chong Li, to name just a few. 
Huawei Technologies' Central Software Institute supports many efforts in parallel programming with the BSP model.


%

%

\bibliographystyle{IEEEtran}
\balance

\bibliography{YouryWCCCE2017} 

\end{document}